\documentclass[letter,
               biblatex,     
               keeplastbox,   
               ]{jacow}
\usepackage{pdfpages,multirow,ragged2e} %
\usepackage{xspace}
\usepackage{xcolor}
\usepackage[USenglish]{babel}

\graphicspath{{pdf_figures/}}

\newcommand{\bmad}{\textit{Bmad}\xspace}
\newcommand{\bfx}{\mathbf{x}}
\newcommand{\bfy}{\mathbf{y}}
\newcommand{\bfs}{\mathbf{s}}
\newcommand{\bfu}{\mathbf{u}}
\newcommand{\bfv}{\mathbf{v}}
\newcommand{\bfr}{\mathbf{r}}

\newcommand{\dxds}{\frac{dx}{ds}}
\newcommand{\dyds}{\frac{dy}{ds}}

\newcommand{\Eq}[1]{Eq.~\ref{#1}}
\newcommand{\Min}{\text{min}}
\newcommand{\Max}{\text{max}}

\begin{document}

\title{Simulating a Positron Converter in Bmad}

\author{John Mastroberti, Indiana University, Bloomington, IN, 47405, USA \\
David Sagan and Jim Shanks, Cornell University, Ithaca, NY, 14850, USA}

\maketitle

\begin{abstract} 
A model to describe the output particle distribution generated
by particles impinging on a planar target has been
developed. This model was developed to simulate positron
production in the Cornell CESR Linac but can be applied to targets using different types
of particles.

To model a specific converter target, the output particle distribution is first
simulated by tracking particles using the Geant4 toolkit
which models the fundamental physics of the conversion process. The Monte Carlo
distribution from Geant4 is fitted to a set of functions and the function
coefficients are saved for use in simulations. Using
the fit functions in a simulation is not only more portable but is also more than an order of magnitude faster than running Geant4.
This model has been successfully incorporated into the \bmad simulation toolkit.

The output position, angular orientation, and momentum distribution is modeled.
Preliminary support for modeling the polarization distribution is also
discussed.
\end{abstract}

\section{Introduction}


A common method of producing positrons for use in an accelerator is through the
use of a positron converter\cite{mikhailichenko02, andreani75}. This is
typically a slab of heavy metal, such as tungsten, located in a Linac which
is bombarded with electrons with energies of order $100$ MeV. The
electrons emit photons via bremsstrahlung which in turn decay to $e^+ e^-$
pairs via the reaction 
\begin{align*} 
e^- + Z \rightarrow e^- + Z + \gamma \rightarrow e^- + Z + e^- + e^+.  
\end{align*} 
After the converter, the electrons and any other particles produced can be filtered off with a dipole
magnet effectively ``converting" a beam of incoming electrons into a beam
of outgoing positrons.

Both bremsstrahlung from electrons traveling through the field of a nucleus
and the subsequent $e^+ e^-$ pair production have been discussed at length in
the literature\cite{tsai66,tseng71,seltzer85,seltzer86,poskus18,poskus19,tsai74}.
Many of the
theoretical treatments of bremsstrahlung and $e^+ e^-$ pair production are
reviewed by Tsai\cite{tsai74}.
Analytic expressions for both the bremsstrahlung radiation
spectrum and secondary electron energy spectrum as functions of target
thickness have been derived\cite{tsai66}, and the problem has been treated
across a wide range of incoming electron energies\cite{tseng71}.  The
bremsstrahlung cross section has been tabulated for a wide range of
electron energies and target materials\cite{seltzer85,seltzer86}.
Recently, computer codes have been developed to calculate the spectra and
angular distributions of bremsstrahlung\cite{poskus18,poskus19}, although
these are designed to work at low energies ($< 3$ MeV).

Despite the extensive study of bremsstrahlung and pair production that exists
in the literature, there is no closed form description of the kinematics of
positrons produced by bremsstrahlung from electrons impinging on a target.
This problem has been treated numerically\cite{b:fromowitz}, but there is no
existing machinery to incorporate these results into a full accelerator
simulation.  These prior efforts also do not address the production of
polarized positrons from a beam of polarized electrons, a topic which is of
current interest\cite{abbott16}.

In response, the authors have developed a converter model and this model 
has been incorporated in the \bmad\cite{b:bmad} accelerator toolkit. The model is designed to be flexible, accommodating any converter material and thickness as well as
supporting arbitrary incoming and outgoing
particle species. The model is also fast to use since the expensive physics calculations are performed
only once when the converter model coefficients are being calculated. This is especially
important when designing a machine with a converter as optimization simulations
are typically time intensive.

One challenge in constructing the model was how to handle the complexity coming from the fact that the outgoing
distribution has five dimensions (two spatial and three momentum) along with the incoming particle momentum, and orientation.
Additionally, the target thickness and particle spin are potential degrees of freedom.

\section{The Converter Model}

The first step is to simulate the conversion process and 
calculate conversion probability coefficients. 
The coefficients can then be used to generate particles
as part of a simulation of the machine the converter is a part of.  

The coefficient calculation has two parts. The first part
is to simulate the quantum electrodynamic\cite{pdg}
interactions of the incoming particles with the converter
to produce bremsstrahlung photons and subsequent positrons. 
The distribution of the outgoing particles calculated in this first step is then used to calculate the coefficients
for the expressions used to model the distribution.

\subsection{Coordinate System}

For concreteness, it will be assumed that the impinging beam is composed of
electrons and the outgoing beam composed of positrons. The model itself is
species agnostic.

\begin{figure}
\centering
\includegraphics[width=0.45\textwidth]{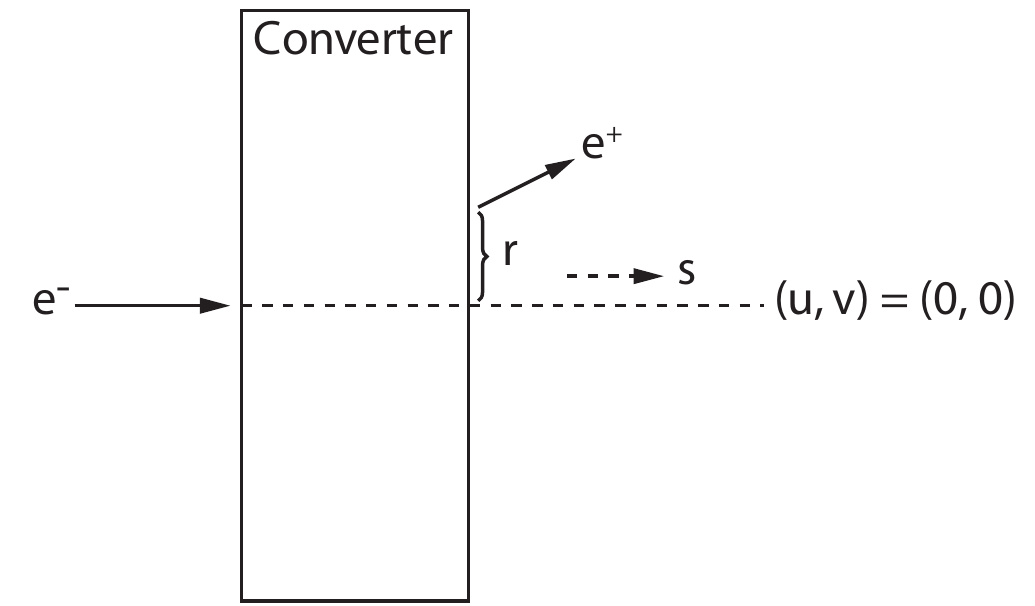}
\caption{Positron converter coordinate system (side view).}
\label{fig:coords1}
\end{figure}

\begin{figure}
\centering
\includegraphics[width=0.4\textwidth]{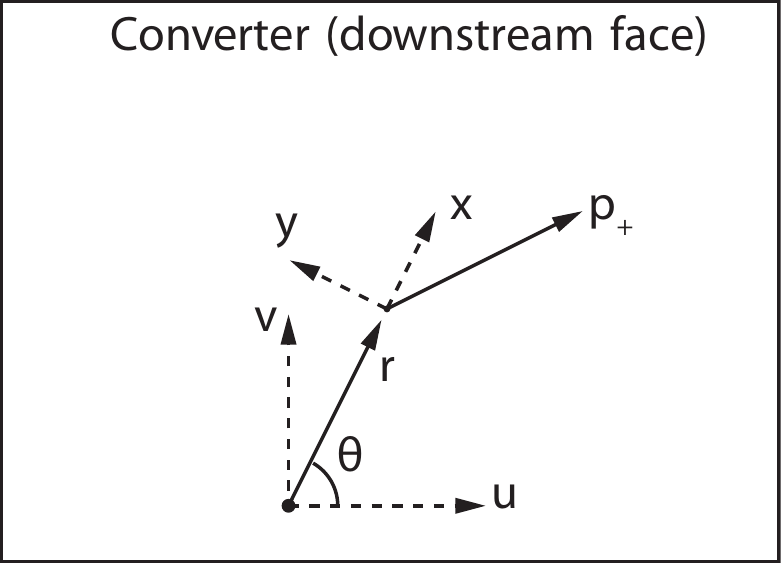}
\caption{Positron converter coordinate system (view of the downstream face).}
\label{fig:coords2}
\end{figure}

Consider an electron incident on the upstream face of a positron converter of
thickness $T$, with momentum $p_-$ perpendicular to the face of the converter
as depicted in Figure \ref{fig:coords1}. The $(\bfu, \bfv, \bfs)$ coordinate system
(Figs. \ref{fig:coords1} and \ref{fig:coords2}) is defined with $\bfs$ perpendicular
to the converter surface and the coordinate origin at the point where the electron
would emerge if it was undeflected through the converter.
Positrons produced in the converter
emerge from the downstream face with some radial displacement $r$ from the origin and at an
angle $\theta$ relative to the $\bfu$ axis as depicted in Figure \ref{fig:coords2}.
Define the rotated ($\bfx$, $\bfy$, $\bfs$) coordinate system so that
$\bfx$ is in the direction of $\bfr$.

By symmetry, $\theta$ must be uniformly distributed from 0 to $2\pi$.
The kinematic properties of the produced positrons which must be
modeled are its radial displacement $r$ and outgoing momentum $\mathbf{p}_+$. The outgoing
momentum is described in terms of its magnitude, $p_+$, and the
slopes along the $\bfx$ and $\bfy$ directions: 
\begin{equation}
  \frac{dx}{ds} = \frac{p_x}{p_s}, \qquad
  \frac{dy}{ds} = \frac{p_y}{p_s}
\end{equation}

\subsection{Probability Distributions}

At the downstream surface, positrons produced in the converter are described by the distribution
\begin{align}
P \left( p_+, r, \dxds, \dyds ; p_-, t \right),
\end{align} 
$P$ is the probability, per incoming electron, of producing an outgoing positron with momentum magnitude
$p_+$, radial displacement $r$, and momentum orientation $(dx/ds, dy/ds)$. $P$ will have a dependence on the converter thickness
$t$ and incoming momentum $p_-$ (non-normal incidence is discussed below). 
The converter thickness dependence is important if the 
converter thickness is allowed to vary in a simulation
where the positron yield is to be maximized. To model the thickness and electron momentum dependence, the distribution $P$ 
is calculated at a number of user defined thicknesses $t$ and
electron momenta $p_-$. Interpolation is then used when generating particles with the model.

For a given $t$ and $p_-$, the integral of $P$ over the space $(p_+, r, dx/ds, dy/ds)$ will be the number of positrons $n_+$
produced per electron
\begin{align} 
  \int P \left( p_+, r, \dxds, \dyds \right)
  \, dp_+ \, dr \, d \! \left( \dxds \right) d \! \left( \dyds \right) & = n_+
\end{align} 
To simplify the computations, $P$ is decomposed into two
subdistributions, 
\begin{align} 
  P \left( p_+, r, \dxds, \dyds \right) & = P_1 \!
  \left( p_+, r \right) \, P_2 \left( \dxds, \dyds ; p_+, r \right) 
\end{align}
where $P_1$ is the probability, per incident electron, of producing a positron with momentum magnitude
$p_+$ and radial displacement $r$, and $P_2$ is the probability, for a given $p_+$ and $r$, of producing
a positron with momentum orientation $(dx/ds, dy/ds)$. $P_1$ is normalized to $n_+$ and $P_2$ is
normalized to one.

The $P_1$ probability distribution is characterized by a two dimensional lookup table that gives $P_1$
at specific values of $p_+$ and $r$. $P_2$ is approximated using a heuristically derived skewed Lorentzian distribution
\begin{align}
  P_2 \left( \dxds, \dyds; p_+, r \right) & = 
  A \frac{1 + \beta \dxds}{1 + \alpha_x^2 \left( \dxds - c_x \right)^2 + \alpha_y^2 \left( \dyds \right)^2} 
  \label{eq:cauchy}
\end{align}
where, as explained below, $\beta$, $\alpha_x$, $\alpha_y$ and $c_x$ will be characterized as functions of $p_+$ and $r$,
and $A$ is calculated from the normalization of $P_2$.

\begin{figure} 
\centering 
\includegraphics[width=0.5\textwidth]{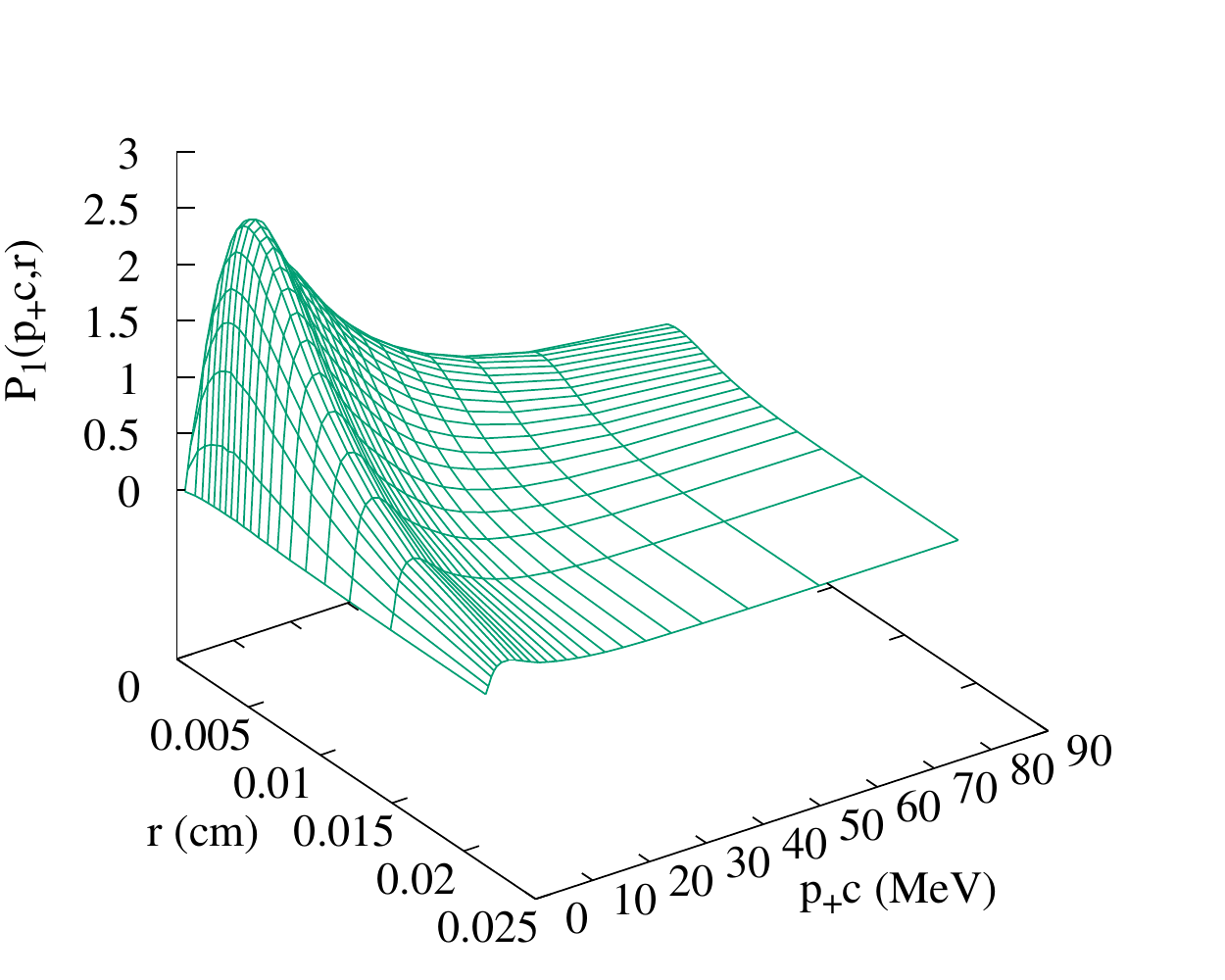}
\caption{$P_1(p_+, r)$ for incoming electrons with $p_- c = 250$ MeV and a
tungsten target of thickness $T = 6.35$ mm.} 
\label{fig:p1} 
\end{figure}

To calculate coefficients, a program to simulate the passage of particles through matter is needed. This
program was developed using the Geant4\cite{geant} particle physics library. A large number of electrons incident upon the
converter are simulated and the produced positron distribution is recorded. 
From this, a fitting program is used to construct a two dimensional binned table of output probability $P_1$ as a function of
$p_+$ and $r$. The user may specify values for $p_+$ and $r$ to bin at or specify
the desired number of bins in both dimensions to use. If the latter option is made, the fitting program will space the bins
to give roughly equal number of particles in each bin. An example
is shown in Figure \ref{fig:p1}. The probability of having a positron
outside of the range of the table is taken to be zero. Inside the range of the table,
the value of $P_1(p_+, r)$ is found by linear interpolation of the table data.

For the $P_2$ modeling, for each of the $(p_+, r)$ bin points for $P_1$, a least squares fit is made using \Eq{eq:cauchy} and
the distribution of $dx/ds$ and $dy/ds$ for particles in the bin. 
Since the distribution in Equation \ref{eq:cauchy} is not normalizable over the entire
$(dx/ds, dy/ds)$ plane, as part of the fit, the fitting program calculates values for $dx/ds_\Min$, $dx/ds_\Max$,
and $|dy/ds|_\Max$, outside of which the value of $P_2$ is taken to be zero. 
These cuts are made so that $P_2$ covers 95\% of the positrons generated by Geant4.
The values of $c_x$, $\alpha_x$, $\alpha_y$, $\beta$, $dx/ds_\Min$, $dx/ds_\Max$,
and $|dy/ds|_\Max$ are obtained by a fit of the Geant4 data in each $(p_+, r)$ bin with
the value of $A$ fixed by the unit normalization of $P_2$.

Once tables for the parameters $c_x$, $\alpha_x$, etc. are calculated at the bin points, the tables are used for each of these parameters to fit the parameters as a function of $p_+$ and $r$. The fit for a given parameter is has two pieces. Above a user definable threshold in $p_+$, a two dimensional fit in $(p_+, r)$ is made of the form:
\begin{equation}
  f(p_+) g(r) \exp(k_1 p_+ + k_2 r) + C
\end{equation}
where $f(p_+)$ and $g(r)$ are third-order polynomials. Below the threshold, for each $p_+$ in the $P_1$ table, a 1D fourth-order polynomial fit in $r$ is made.
The 1D fits are performed for small values of $p_+$ because equation 6 tends to fit the Geant4 data poorly when $p_+ \lesssim 10 \mbox{ MeV}$.
Since most produced positrons are produced in this region, it is desirable for the $P_2$ parameter fits to be more detailed when $p_+$ is small.

\begin{figure} 
\centering 
\includegraphics[width=0.45\textwidth]{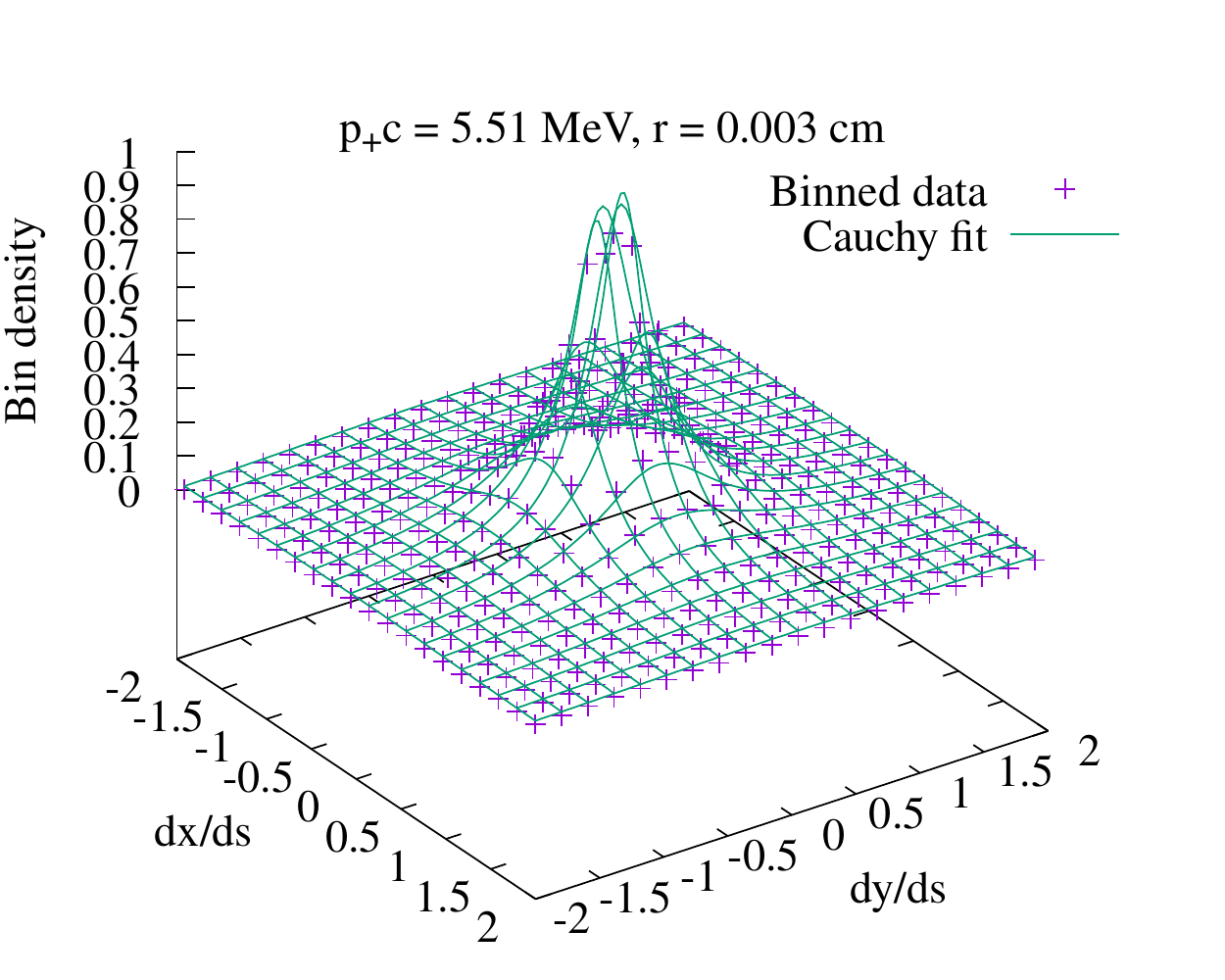}
\caption{$P_2 \left( \dxds, \dyds ; p_+, r \right)$ for incoming electrons
with $p_- c = 250$ MeV and a tungsten target of thickness $T = 6.35$ mm, and
outgoing positrons with $p_+ c = 5.15$ MeV and $r = 0.37$ mm.  The purple
points indicate data obtained directly from the Geant simulation, while the
green curve shows the fit to the data.} 
\label{fig:p2} 
\end{figure}

Examples of the $P_1$ and $P_2$ distributions obtained from the
Geant simulation are shown in Figures \ref{fig:p1} and \ref{fig:p2}
respectively.
The fit given in equation \ref{eq:cauchy} models the data from Geant quite successfully, both when the parameters are obtained by direct fit to the data and when the parameters from the second order fits are used.
The average $\chi^2$ value for the direct fits is about 0.11, while the average $\chi^2$ for the second order fits is about 0.14.
The bins where the fit performs the worst have a $\chi^2$ of about 1 for the direct fits, while the maximum $\chi^2$ for the second order fits is about 2.2.

\begin{figure}
\centering
\includegraphics[width=0.45\textwidth]{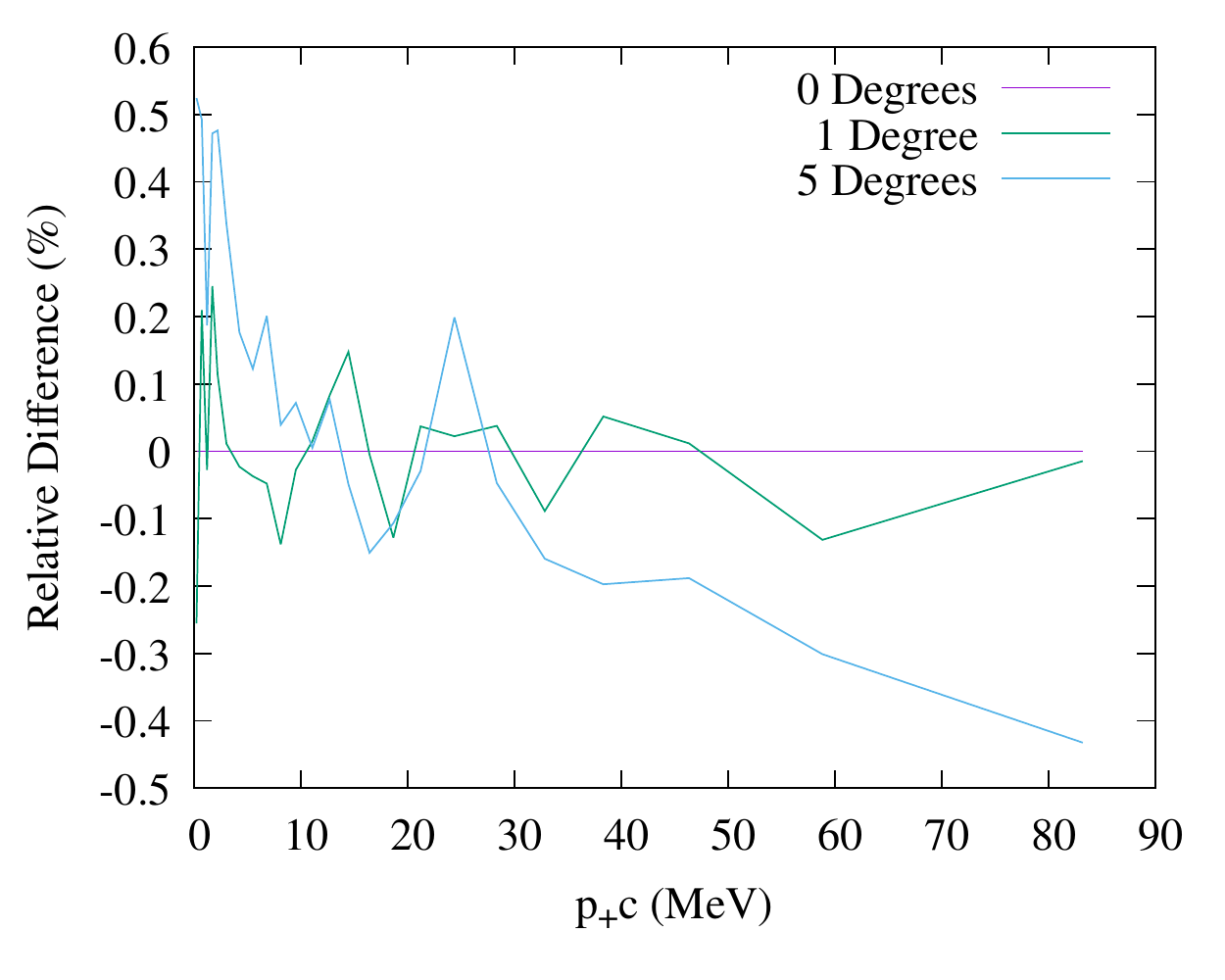}
\caption{
Relative discrepancy of the distributions of positron momenta 
for $\phi = 1$ and 5 degrees compared to the distribution for $\phi = 0$.
}
\label{fig:e_angle}
\end{figure}

\begin{figure}
\centering
\includegraphics[width=0.45\textwidth]{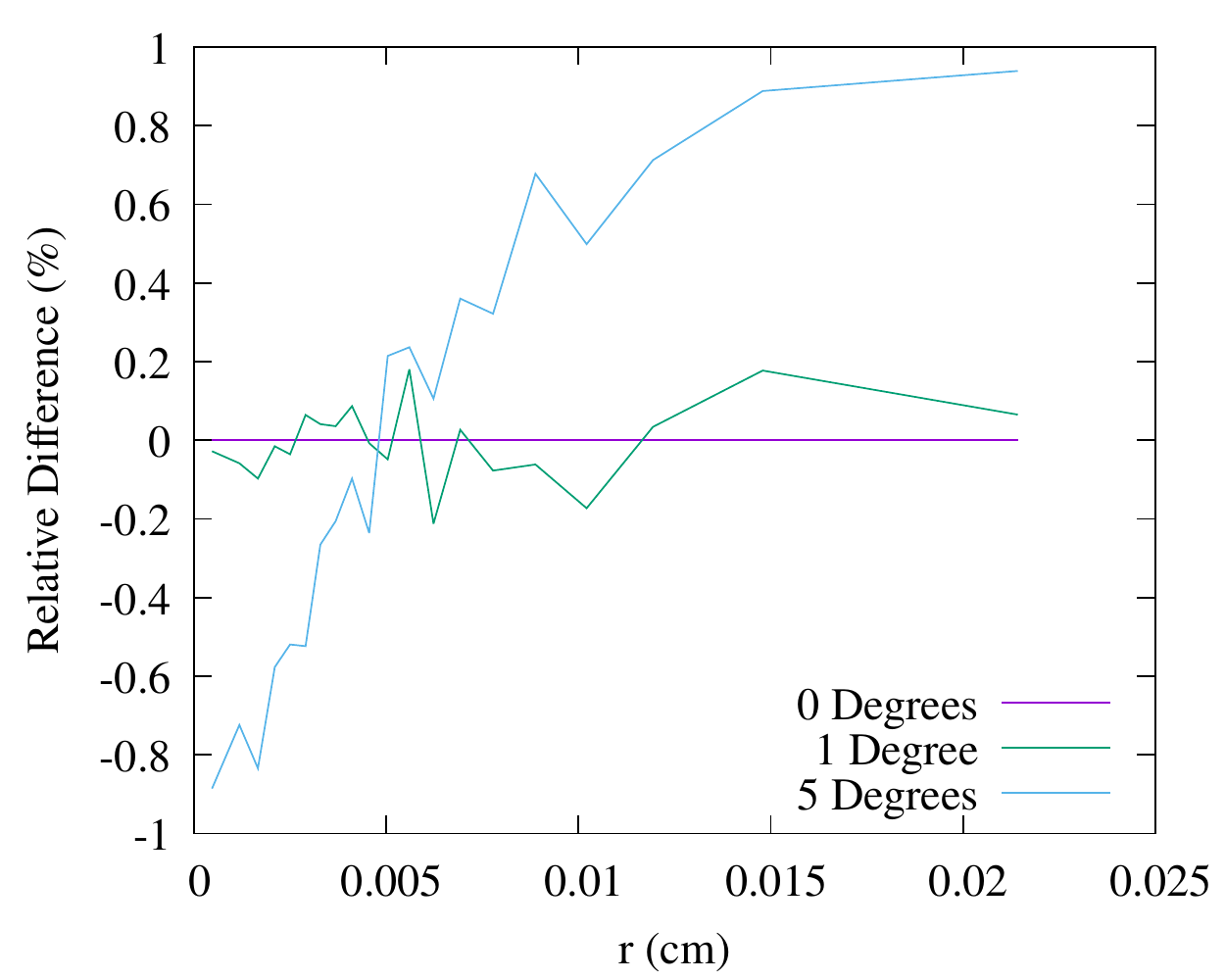}
\caption{
Relative discrepancy of the distributions of positron radial displacements $r$
for $\phi = 1$ and 5 degrees compared to the distribution for $\phi = 0$.
}
\label{fig:r_angle}
\end{figure}

\begin{figure}
\centering
\includegraphics[width=0.45\textwidth]{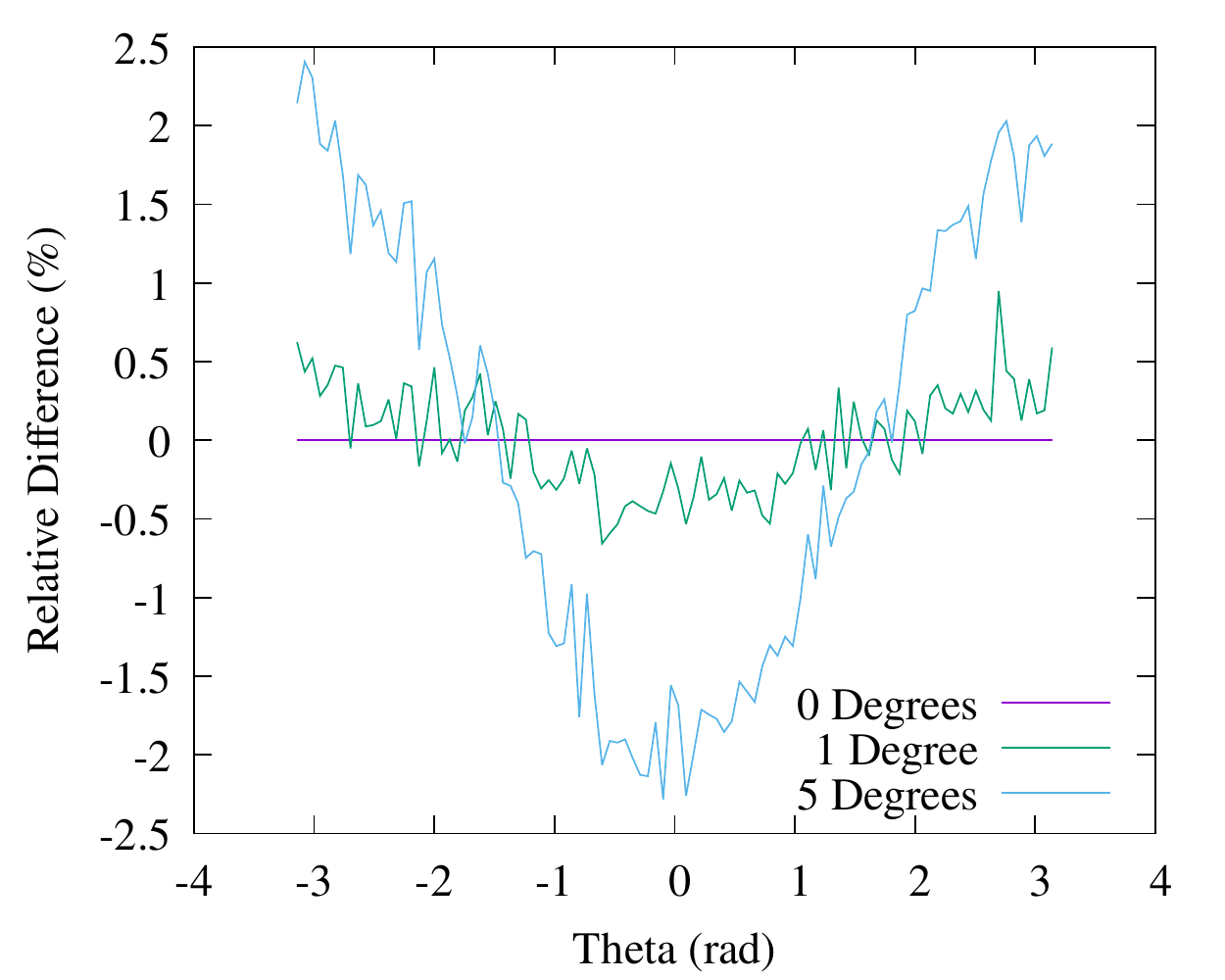}
\caption{
Relative discrepancy of the distributions of positron angle $\theta$
for $\phi = 1$ and 5 degrees compared to the distribution for $\phi = 0$.
}
\label{fig:theta_angle}
\end{figure}

The Geant4 simulation and accompanying formalism assume that the incoming electrons
are normally incident on the target. To adjust for a non-normal impinging electron when
generating positrons with the converter model,
the coordinate system is rotated so that the $\bfs$ axis is aligned along the incoming electron axis.  
The effective converter thickness used is adjusted to be $T \sec\phi$ where $\phi$ is the angle from the normal, as this is the
distance from the electron beam's entry point on the converter's upstream face
to it's exit point on the downstream face.
The distributions of positron momentum $p_+$, radial displacement $r$, and outgoing angle $\theta$ are shown in Figures \ref{fig:e_angle}-\ref{fig:theta_angle} for various incoming angles $\phi$. Up to 5 degrees off-normal 
there is very little deviation from the on-axis results. For a machine like the Cornell CESR Linac, where the impinging beam has an
angular spread well less than 1 degree, the simple treatment of off-normal is a good approximation.

\section{Spin Tracking}

Polarization transfer from the incoming electrons to the outgoing positrons has
also been modeled.  Any incoming polarization $\mathbf{S}_-$ may be specified,
and histograms describing $S_x$, $S_y$, and $S_z$ of the produced positrons as
functions of $p_+$ and $r$ are produced. In the Geant4 simulations done for this paper,
it is observed that only
the longitudinal polarization of the incoming electrons is ever transferred to
the produced positrons; the produced positrons always have $S_x$ and $S_y$
essentially zero regardless of the incoming electron polarization.  


Recently, the PEPPo collaboration\cite{abbott16} has published experimental
results of polarized positron production from a polarized electron beam via
bremsstrahlung, to which the spin tracking results from this simulation can be
compared.  This simulation yields polarization transfer efficiencies
significantly higher than those observed experimentally for outgoing positrons
with $p_+ c$ less than 5 MeV.  In contrast, the simulation's results for higher
momentum positrons agrees closely with experiment. The discrepancy is likely
due to approximations made by the Geant4 library in computing polarization
transfer during the pair production step. While this disagreement shows that the
spin tracking components of this simulation should not be relied on for
accuracy at present, all of the infrastructure is in place to handle spin
tracking.  When more accurate spin tracking methods become available in Geant,
this simulation, and the \bmad converter element, will be ready to use their
results.
\section{The Bmad Converter Element}

\bmad is a software toolkit for the simulation of high energy charged particles
and X-rays. To make use of the converter model,
a converter element has been added to \bmad which uses the model coefficients
as described above allowing
for such things as a start to end simulation of the Cornell CESR Linac.
The model coefficients are stored within a lattice input file which makes 
converter simulation portable to any \bmad based program.

For each incoming electron, \bmad randomly pulls values of $p_+$, $r$,
$\dxds$, and $\dyds$ from the interpolated probability distributions. 
Outgoing positrons are assigned a weighting factor proportional
to the likelihood of producing a positron from the given incoming electron. 
This weighting factor is then used with any statistical analysis.

\begin{figure} 
\centering 
\includegraphics[width=0.45\textwidth]{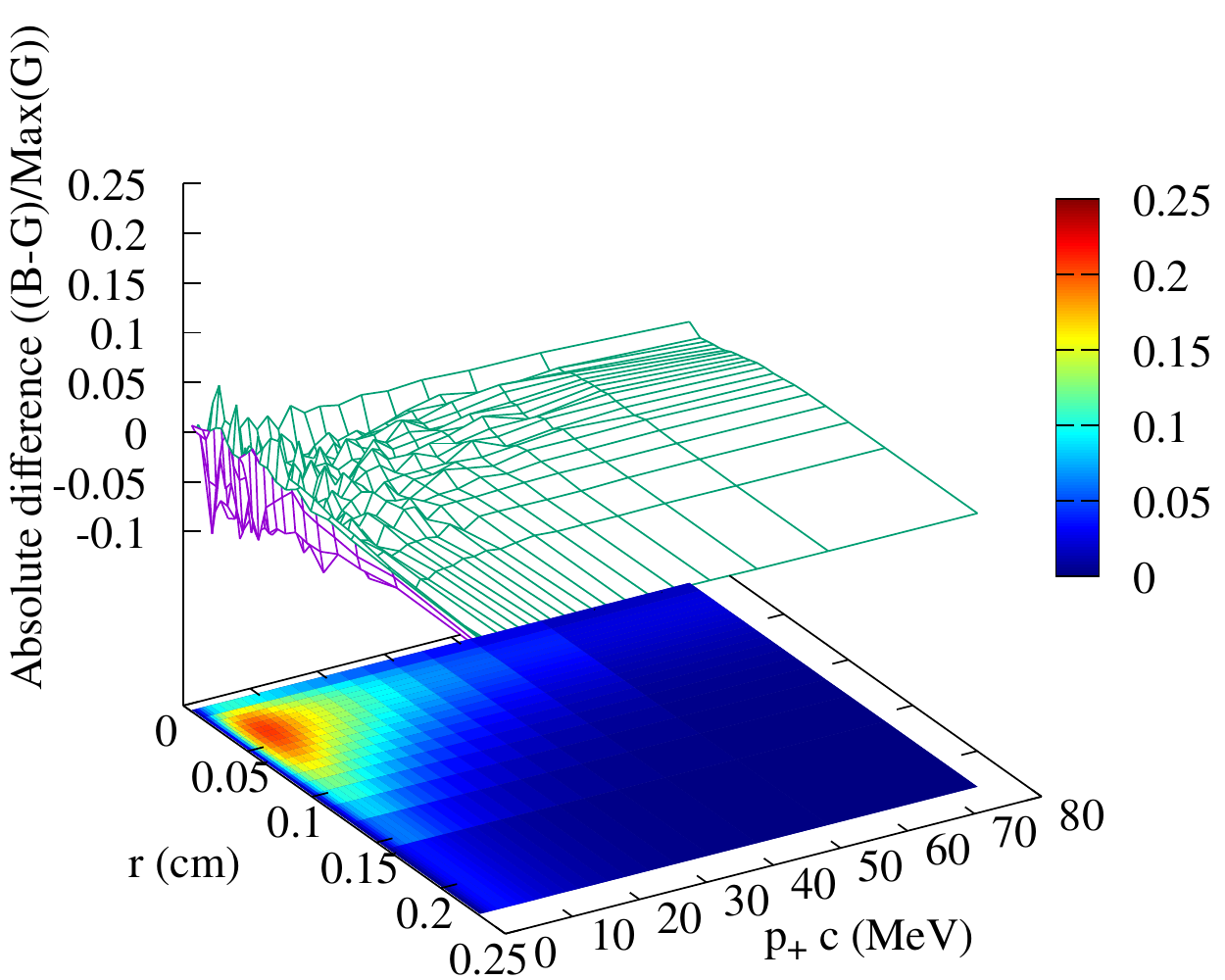}
\caption{Comparison of the \bmad converter element with the Geant simulation.
The surface plot shows the difference between the $P_1$ distributions generated by \bmad and by Geant, with the difference scaled by the maximum value of the $P_1$ distribution from Geant.
The color plot shows the value of the $P_1$ distribution from Geant, indicating the region where obtaining accurate results from \bmad is most important.}
\label{fig:bmad} 
\end{figure}

Agreement between direct output from Geant4 and emulated output from \bmad
is quite good, with the distribution of particles from \bmad coming within a
few percent of the Geant distributions.  This is illustrated in Figure
\ref{fig:bmad}. \bmad simulates the positron converter
approximately 20 times faster than the direct Geant4 physics simulations.

As the figure shows, the relative discrepancy between the distributions
of positrons produced by Geant and \bmad agree within a few percent under most circumstances.
The two scenarios where this is not the case are for the lowest energy bins, and for bins
with large $p_+$ and large $r$. The discrepancy in the first case is due to the fact that \bmad
does not produce positrons with $p_+$ below the value of $p_+$ used as the center of the first bin.
This causes \bmad to produce less low-energy positrons than Geant. This discrepancy is acceptable,
as virtually all of these low-energy positrons would be lost within a short distance from the converter.
The discrepancy for large $p_+$ and $r$ bins is also acceptable since very few positrons are produced
in this region.

%
%
%

%
%
%
%
%
%

\section{Conclusion}

The converter model presented in this paper, when applied to positron production,
provides simulation results that
are in close agreement to Geant4's direct simulation of bremsstrahlung
radiation and pair production.  A small sacrifice in accuracy is made in return
for a significant increase in simulation speed, making this method appropriate
for use in lattice simulation and optimization software such as \bmad.

One of the primary motivators for this work was the desire to fully model the
CESR linac lattice.  Work is currently underway to use the model
developed here. A first goal in this endeavor is to assess the agreement
between the \bmad model of the linac as a whole the against the linac's experimentally
observed beam.  Once good agreement is established, \bmad can be
used to optimize the lattice design, including quadrupole and solenoid tuning.
Other future applications could include the optimization of element positioning
on the linac beam line.

\section{Acknowledgments}

Many thanks to Vardan Khachatryan for helping with running of Geant.
This work is supported by National Science Foundation award number 
DMR-1829070.


\printbibliography

@PHDTHESIS{b:fromowitz,
       author = {{Fromowitz}, Daniel Bret},
        title = "{Increasing the positron capture efficiency of the CESR linac injector}",
       school = {Cornell University},
         year = 2000,
        month = Oct
}

@article{geant,
title = "Geant4—a simulation toolkit",
journal = "Nuclear Instruments and Methods in Physics Research Section A: Accelerators, Spectrometers, Detectors and Associated Equipment",
volume = "506",
number = "3",
pages = "250 - 303",
year = "2003",
issn = "0168-9002",
doi = "https://doi.org/10.1016/S0168-9002(03)01368-8",
url = "http://www.sciencedirect.com/science/article/pii/S0168900203013688",
author = "S. Agostinelli and J. Allison and K. Amako and J. Apostolakis and H. Araujo and P. Arce and M. Asai and D. Axen and S. Banerjee and G. Barrand and F. Behner and L. Bellagamba and J. Boudreau and L. Broglia and A. Brunengo and H. Burkhardt and S. Chauvie and J. Chuma and R. Chytracek and G. Cooperman and G. Cosmo and P. Degtyarenko and A. Dell'Acqua and G. Depaola and D. Dietrich and R. Enami and A. Feliciello and C. Ferguson and H. Fesefeldt and G. Folger and F. Foppiano and A. Forti and S. Garelli and S. Giani and R. Giannitrapani and D. Gibin and J.J. Gómez Cadenas and I. González and G. Gracia Abril and G. Greeniaus and W. Greiner and V. Grichine and A. Grossheim and S. Guatelli and P. Gumplinger and R. Hamatsu and K. Hashimoto and H. Hasui and A. Heikkinen and A. Howard and V. Ivanchenko and A. Johnson and F.W. Jones and J. Kallenbach and N. Kanaya and M. Kawabata and Y. Kawabata and M. Kawaguti and S. Kelner and P. Kent and A. Kimura and T. Kodama and R. Kokoulin and M. Kossov and H. Kurashige and E. Lamanna and T. Lampén and V. Lara and V. Lefebure and F. Lei and M. Liendl and W. Lockman and F. Longo and S. Magni and M. Maire and E. Medernach and K. Minamimoto and P. Mora de Freitas and Y. Morita and K. Murakami and M. Nagamatu and R. Nartallo and P. Nieminen and T. Nishimura and K. Ohtsubo and M. Okamura and S. O'Neale and Y. Oohata and K. Paech and J. Perl and A. Pfeiffer and M.G. Pia and F. Ranjard and A. Rybin and S. Sadilov and E. Di Salvo and G. Santin and T. Sasaki and N. Savvas and Y. Sawada and S. Scherer and S. Sei and V. Sirotenko and D. Smith and N. Starkov and H. Stoecker and J. Sulkimo and M. Takahata and S. Tanaka and E. Tcherniaev and E. Safai Tehrani and M. Tropeano and P. Truscott and H. Uno and L. Urban and P. Urban and M. Verderi and A. Walkden and W. Wander and H. Weber and J.P. Wellisch and T. Wenaus and D.C. Williams and D. Wright and T. Yamada and H. Yoshida and D. Zschiesche"
}

@article{b:bmad,
  author  = "D. Sagan",
  title   = "Bmad: A relativistic charged particle simulation library",
  journal = "Nuc. Instrum. and Methods in Phys Research A",
  volume  = "558", 
  pages   = "356-359", 
  year    = "2006",
  note    = "\url{http://www.lepp.cornell.edu/~dcs/bmad}"
}

@article{abbott16,
  title = {Production of Highly Polarized Positrons Using Polarized Electrons at MeV Energies},
  author = {Abbott, D. and Adderley, P. and Adeyemi, A. and Aguilera, P. and Ali, M. and Areti, H. and Baylac, M. and Benesch, J. and Bosson, G. and Cade, B. and Camsonne, A. and Cardman, L. S. and Clark, J. and Cole, P. and Covert, S. and Cuevas, C. and Dadoun, O. and Dale, D. and Dong, H. and Dumas, J. and Fanchini, E. and Forest, T. and Forman, E. and Freyberger, A. and Froidefond, E. and Golge, S. and Grames, J. and Gu\`eye, P. and Hansknecht, J. and Harrell, P. and Hoskins, J. and Hyde, C. and Josey, B. and Kazimi, R. and Kim, Y. and Machie, D. and Mahoney, K. and Mammei, R. and Marton, M. and McCarter, J. and McCaughan, M. and McHugh, M. and McNulty, D. and Mesick, K. E. and Michaelides, T. and Michaels, R. and Moffit, B. and Moser, D. and Mu\~noz Camacho, C. and Muraz, J.-F. and Opper, A. and Poelker, M. and R\'eal, J.-S. and Richardson, L. and Setiniyaz, S. and Stutzman, M. and Suleiman, R. and Tennant, C. and Tsai, C. and Turner, D. and Ungaro, M. and Variola, A. and Voutier, E. and Wang, Y. and Zhang, Y.},
  collaboration = {PEPPo Collaboration},
  journal = {Phys. Rev. Lett.},
  volume = {116},
  issue = {21},
  pages = {214801},
  numpages = {5},
  year = {2016},
  month = {May},
  publisher = {American Physical Society},
  doi = {10.1103/PhysRevLett.116.214801},
  url = {https://link.aps.org/doi/10.1103/PhysRevLett.116.214801}
}

@article{tsai66,
  title = {Thick-Target Bremsstrahlung and Target Considerations for Secondary-Particle Production by Electrons},
  author = {Tsai, Y. S. and Van Whitis},
  journal = {Phys. Rev.},
  volume = {149},
  issue = {4},
  pages = {1248--1257},
  numpages = {0},
  year = {1966},
  month = {Sep},
  publisher = {American Physical Society},
  doi = {10.1103/PhysRev.149.1248},
  url = {https://link.aps.org/doi/10.1103/PhysRev.149.1248}
}

@article{poskus18,
title = {BREMS: A program for calculating spectra and angular distributions of bremsstrahlung at electron energies less than 3 MeV},
journal = {Computer Physics Communications},
volume = {232},
pages = {237-255},
year = {2018},
issn = {0010-4655},
doi = {https://doi.org/10.1016/j.cpc.2018.04.030},
url = {https://www.sciencedirect.com/science/article/pii/S0010465518301462},
author = {Andrius Poškus},
}

@article{poskus19,
title = {Shape functions and singly differential cross sections of bremsstrahlung at electron energies from 10 eV to 3 MeV for Z = 1–100},
journal = {Atomic Data and Nuclear Data Tables},
volume = {129-130},
pages = {101277},
year = {2019},
issn = {0092-640X},
doi = {https://doi.org/10.1016/j.adt.2019.03.002},
url = {https://www.sciencedirect.com/science/article/pii/S0092640X19300099},
author = {Andrius Poškus},
}

@article{seltzer85,
title = {Bremsstrahlung spectra from electron interactions with screened atomic nuclei and orbital electrons},
journal = {Nuclear Instruments and Methods in Physics Research Section B: Beam Interactions with Materials and Atoms},
volume = {12},
number = {1},
pages = {95-134},
year = {1985},
issn = {0168-583X},
doi = {https://doi.org/10.1016/0168-583X(85)90707-4},
url = {https://www.sciencedirect.com/science/article/pii/0168583X85907074},
author = {Stephen M. Seltzer and Martin J. Berger},
}

@article{seltzer86,
title = {Bremsstrahlung energy spectra from electrons with kinetic energy 1 keV–10 GeV incident on screened nuclei and orbital electrons of neutral atoms with Z = 1–100},
journal = {Atomic Data and Nuclear Data Tables},
volume = {35},
number = {3},
pages = {345-418},
year = {1986},
issn = {0092-640X},
doi = {https://doi.org/10.1016/0092-640X(86)90014-8},
url = {https://www.sciencedirect.com/science/article/pii/0092640X86900148},
author = {Stephen M. Seltzer and Martin J. Berger},
}

@article{tsai74,
  title = {Pair production and bremsstrahlung of charged leptons},
  author = {Tsai, Yung Su},
  journal = {Rev. Mod. Phys.},
  volume = {46},
  issue = {4},
  pages = {815--851},
  numpages = {0},
  year = {1974},
  month = {Oct},
  publisher = {American Physical Society},
  doi = {10.1103/RevModPhys.46.815},
  url = {https://link.aps.org/doi/10.1103/RevModPhys.46.815}
}

@article{tseng71,
  title = {Exact Screened Calculations of Atomic-Field Bremsstrahlung},
  author = {Tseng, H. K. and Pratt, R. H.},
  journal = {Phys. Rev. A},
  volume = {3},
  issue = {1},
  pages = {100--115},
  numpages = {0},
  year = {1971},
  month = {Jan},
  publisher = {American Physical Society},
  doi = {10.1103/PhysRevA.3.100},
  url = {https://link.aps.org/doi/10.1103/PhysRevA.3.100}
}

@inproceedings{mikhailichenko02,
  title = "{CESR's Positron Source}",
  author = {Mikhailichenko, A.},
  booktitle = {SLAC LC02},
  year = {2002},
  month = {Feb}
}

@article{andreani75,
title = {Positron converter for the Frascati linear accelerator},
journal = {Nuclear Instruments and Methods},
volume = {129},
number = {2},
pages = {365-371},
year = {1975},
issn = {0029-554X},
doi = {https://doi.org/10.1016/0029-554X(75)90726-0},
url = {https://www.sciencedirect.com/science/article/pii/0029554X75907260},
author = {R. Andreani and A. Cattoni}
}

@article{pdg,
    author = "Zyla, P.A. and others",
    collaboration = "Particle Data Group",
    title = "{Review of Particle Physics}",
    doi = "10.1093/ptep/ptaa104",
    journal = "PTEP",
    volume = "2020",
    number = "8, sec. 34.4",
    pages = "083C01",
    year = "2020"
}

\clearpage

\end{document}